\documentclass[12pt,a4paper]{article}

\begin{document}

\title{\textbf{Principles of Discrete Time Mechanics:}\\
$\mathbf{III}$\textbf{. Quantum Field Theory}}
\author{Keith Norton and George Jaroszkiewicz$^{*}$ \\
%EndAName
$^{*}$Department of Mathematics, University of Nottingham\\
University Park, Nottingham NG7 2RD, UK}
\date{\today }
\maketitle

\begin{abstract}
\textit{We apply the principles discussed in earlier papers to the
construction of discrete time quantum field theories. We discuss some of the
issues to do with loss of Lorentz covariance and its recovery in the
appropriate limit. We use the Schwinger action principle to find the
discrete time free field commutators for scalar fields, which allows us to
set up the reduction formalism for discrete time scattering processes. Then
we derive the discrete time analogue of the Feynman rules for a scalar field
with a cubic self interaction and give examples of discrete time scattering
amplitude calculations. We find overall conservation of total linear
momentum and overall conservation of total }$\theta $ \textit{parameters,
which is the discrete time analogue of energy conservation and corresponds
to the existence of a Logan invariant for the system. We find that temporal
discretisation leads to softened vertex factors, modifies propagators and
gives a natural cutoff for physical particle momenta.}
\end{abstract}

\section{Introduction}

THIS paper is the third in a series devoted to the construction of discrete
time classical and quantum mechanics, based on the notion that there is a
fundamental interval of time, $T$. The objective is to investigate the
properties of a dynamics where continuity in time, and hence
differentiability with respect to time, has been abolished. With no
velocities, there are no Lagrangians in the ordinary sense, and then there
are no canonical conjugate momenta or Hamiltonians either. It would appear
then to be a catastrophic recipe for recasting the laws of classical and
quantum physics, but as we try to show in this paper, this is not really the
case. Moreover, there is every prospect for finding novel features of the
dynamics not encountered in continuous time mechanics which may go some way
towards alleviating the divergence problems encountered in conventional
quantum field theory.

The first paper of this series, referred to as \emph{Paper} $I$ $\cite{J&N-I}%
,$ introduced basic principles for the temporal discretisation of continuous
time classical and quantum particle mechanics. The second paper, referred to
as \emph{Paper} $II$ $\cite{J&N-II},$ applied these principles to classical
field theory, including gauge invariant electrodynamics and the Dirac field.
These papers should be consulted for further explanation of our notation and
methodology. In this paper, referred to as \emph{Paper} $III$, we apply the
techniques of \emph{Paper} $I$ to the quantisation of the scalar field
systems studied in \emph{Paper} $II$, i.e., we discuss quantised discrete
time scalar field theory.

Following the analysis of the earlier papers, we denote by $\mathcal{D}$ our
process of discretising time using virtual paths and by $\mathcal{Q}$ the
process of quantisation using transition amplitudes based on the system
function, each of these processes being applied to some classical Lagrangian 
$L$. Then we can say that \emph{Paper} $I$ discusses models of type $%
\mathcal{D}L$ and $\mathcal{QD}L$ whereas \emph{Paper} $II$ discusses models
of type $\mathcal{DL}$, where $\mathcal{L}$ is a Lagrange density. Now since
such a density may be associated with the first quantisation of a classical
theory, i.e., to $\mathcal{Q}L$ models, as discussed in \emph{Paper} $II$
for the Schr\"{o}dinger equation, models of type $\mathcal{DL}$ may be
regarded as equivalent in some sense to those of type $\mathcal{DQ}L.$ This
allows a direct comparison of the processes $\mathcal{DQ}$ and $\mathcal{QD}$%
, and in \emph{Paper} $II$ it was argued that these are not the same in
general.

The present paper considers models of type $\mathcal{QDL}$. Because such
models may be regarded as equivalent in the above sense to those of type $%
\mathcal{QDQ}L,$ then \emph{Paper} $III$ may be considered to be a
discussion of discrete time second quantisation. Note however that the $%
\mathcal{QDQ}$ process used in this paper is not in general equivalent to
the process $\mathcal{DQQ}$ because the $\mathcal{D}$ and $\mathcal{Q\,}$
processes do not commute. This means that our paper discusses the
quantisation of discrete time classical field theories and not the temporal
discretisation of quantum field theories, such as in lattice gauge theories.
In the latter, discretisation is regarded as an approximation which becomes
exact in the continuum limit. In our approach our mechanics is regarded as
exact at each stage and the continuum limit is taken only to make
comparisons with standard formulations. This is a significant difference
between our approach and various other formulations using a discrete time,
because of our insistence on adhering to the principles of the formulation
at all stages. In particular, the constants of the motion are constructed to
be exact and not approximately conserved up to some powers of $T.$

An important question which arises naturally in the context of discrete time
and/or space mechanics is that of Lorentz covariance. Our answer is that
Lorentz symmetry of say scattering matrix elements emerges in the
appropriate limit, such as $T\rightarrow 0$, and other than that, is not
really something to worry about, as it is regarded here as an approximation
to a deeper underlying structure. An analogy with representational art is
useful here. If we liken continuous time theories to pictures drawn on
normal canvas, then our discrete time mechanics is a picture drawn on a
conventional analogue television screen. In the former model of spacetime it
is frequently speculated that continuity might break down, perhaps at Planck
scales (we do know that a real canvas is made up of atoms), but otherwise,
continuity on the plane of the canvas exists at all levels and carries with
it all the associated symmetries of the plane, such as translation and
rotational invariance. On a television screen, however, we have two
perspectives. From a distance, a television picture really does look like
one painted on a canvas, but a closer look would readily show the horizontal
lines which make up the picture. There is a discreteness vertically, but a
continuity horizontally. Likewise, in discrete time mechanics, there is a
discreteness along the time axis with all the normal continuity along the
space axis. Like a television picture, there is less symmetry when viewed
close up than when viewed at a distance, and it would be futile and in
principle wrong to try to pretend that such long-distance symmetries should
exist at all scales. What we are doing, therefore, is more like exploring
the mechanics of a television set rather than the pictures drawn on it. This
suggests that discretisation of time in the context of General Relativity is
an obvious candidate for investigation.

The art analogy can be pursued further. Discretisation of space as well as
time, such as in lattice gauge theories and the work of authors such as
Bender et al \cite{BENDER} and Yamamoto et al \cite{YAMAMOTO.95}, gives a
lattice space-time picture which corresponds to what occurs on a computer
monitor, where the picture is fully digitised. This form of discrete
space-time mechanics is inherently different to our discrete time mechanics
and the two should not be confused.

Another important question related to the issue of Lorentz covariance is: 
\textit{in which inertial frame are we discretising time?} Of course, if we
believed in an absolute time in the strict sense of Newton then we would
have an immediate answer. However, we are approaching discrete time from a
more modern perspective and the problem is a very real one for us. Our
answer is to go beyond special relativity and consider cosmological
perspectives. It is an empirical fact that, from the point of view of
observers on the earth, we are moving at a speed of about $500$ to $600$%
\textit{\ }km per sec relative to a frame of reference in which the cosmic
background radiation field of Penzias and Wilson is isotropic \cite{GUTH}
(the so-called \textit{dipole effect}). According to the Cosmological
Principle, we should be able to find a local inertial frame in the
neighbourhood of each point in space time with the same property, i.e. one
in which the cosmic background radiation field is isotropic to a very high
degree, except for tiny ripples equivalent to those recently observed by
COBE \cite{COBE}. This frame should be unique at each point, up to spatial
rotations. We will refer to this frame as the \textit{local absolute frame.}

If we allow that our universe is reasonably well described via a
Robertson-Walker metric, then as we change position and time we expect the
local absolute frame to change as well. However, it will always be
empirically identifiable at each place and time in the universe. The
standard co-ordinate time in such a frame is called the proper or co-moving
time in the usual formulation of Robertson-Walker cosmology, and represents
the proper time of a point particle (or galaxy) at rest relative to the
local cosmic mass distribution. In answer to the question posed above, we
suggest that time is discretised via local absolute frames.

Care should be taken to keep in mind that throughout this paper, when we
discuss Minkowski spacetime and its temporal discretisation, we are really
referring to local inertial frames. Of course, there is the additional
question of local variations due to gravitational disturbances arising from
locally inhomogeneous matter densities. Answering this question amounts to
constructing a discrete time analogue of general relativity, which will be
reserved for a subsequent paper in this series. Since we are interested in
applications to particle theory in this paper, we will not consider these
issues further here, except to make a final observation about this line of
thought. If we were discretising space as well as time (which we are not),
we would have to consider the additional question: \textit{how are we
discretising space? }If we were choosing the simplest sort of discretisation
scheme,\textit{\ }a cubic lattice (say), then we would have to specify three
spatial orthogonal cartesian axes. Until recently there was no evidence of
any spatial anisotropy on truly cosmological scales, so we had no criterion
for picking out any special directions in space. We note however the very
recent observation of the so-called \textit{corkscrew effect} reported by
Nodland and Ralston \cite{NODLAND & RALSTON.97A}, which if confirmed will
require a re-assessment of the position. The comic background radiation
field, however, \textit{does} give us a working prescription for picking out
a unique timelike direction at each point.

Because of the relatively greater complexity of discrete time field theory
compared with conventional field theory, we have restricted our attention in
this paper to scalar fields. The general features found here should find
their direct analogues with the Dirac and Maxwell fields. We reserve the
further discussion of these fields to the next papers in this series. Our
principal aim in this paper is to discuss how the process of discretising
time alters Feynman rules for scattering amplitudes and scattering
cross-sections. Issues of renormalisation are left for later papers in the
series. An important feature of the present investigation is the discrete
time oscillator, which is directly related to free particle states used to
define \textit{in} and \textit{out} states.

In $\S 2$ we discuss the quantisation of scalar fields, using Schwinger's
action principle to derive ground state expectation values of time ordered
products. Then in $\S 3$ we apply these methods to the free neutral scalar
field. The results are in agreement with the more direct calculation of the
quantised discrete time harmonic oscillator discussed in \emph{Paper} $I$.
We examine in more detail the free scalar field propagator and the free
field commutators, the results being consistent with the vacuum expectation
values discussed previously. We find that the free particle creation and
annihilation operators have a natural cutoff in physical particle state
momenta, corresponding to what we call the \textit{elliptic} regime.
Although it is possible to construct linear invariants of the motion outside
this regime, states associated with such operators have various expectation
values which diverge or tend to zero in the infinite time limit, and this
makes them unsuitable for representing physical particles. If $T$ is the
fundamental time interval and the discrete time analogue of energy $E$ is
defined by $E=\sqrt{\mathbf{p.p}+m^2}$ in natural units (where $c=\hbar =1)$%
, then our formulation leads to the condition $TE<\sqrt{12}$ for
physical \textit{in} or \textit{out }particles, which we call the \textit{%
parabolic barrier.} This barrier manifests itself in a number of ways. For
example, the particle flux density associated with each creation operator is
found to be modified by a factor $\sqrt{1-T^2E^2/12}$, which makes physical
sense only in the elliptic regime.

We then turn to interacting scalar fields theories. In $\S 4$ we set up the
discrete time reduction formulae needed to calculate scattering cross
sections and then discuss the perturbative expansion of the vacuum
expectation values of discrete time ordered products for a specific example, 
$\varphi ^3$ scalar field theory. We give the discrete time analogues of the
Feynman rules in configuration space and in momentum space. In $\S 5$ we
present a scattering calculation for the box diagram to illustrate the
formalism and then give general rules for scattering amplitudes. Finally in $%
\S 6$ we give a number of applications of our scattering amplitude rules. We
find that in each case there is a conserved quantity in scattering processes
analogous to energy, related to the existence of what we call a Logan
invariant of the system function. Fortunately, the LSZ formalism is powerful
enough to reveal the existence of such a Logan invariant in a scattering
process without the need for us to find it explicitly for the fully
interacting system.

Our analysis reveals that for $\varphi ^3$ interactions our discrete time
Feynman rules involve vertex softening in the basic diagrams, before any
renormalisation effects are considered. This may be a significant feature of
more realistic interactions. Also, the propagators associated with internal
lines are modified and we use them to show how Lorentz covariance can emerge
as an approximate symmetry of the mechanics. There is therefore some
prospect of our programme making some progress towards the alleviation, if
not complete removal, of divergences in the traditional renormalisation
programme of continuous time relativistic quantum field theory.

\section{The discrete time quantised scalar field}

We turn now to the quantisation of the neutral scalar field. Following the
methodology and notation discussed in \emph{Papers} $I$ and $II,$
particularly the discussion in \emph{Paper} $I$ on the quantised
inhomogeneous oscillator, the discrete time system function for a system
with a scalar field $\varphi $ degree of freedom coupled to a source $j$ is
chosen to be given by 
\begin{equation}
F^n\left[ j\right] =F^n+\frac{{}_1}{{}^2}T\int d^3\mathbf{x}\left\{
j_{n+1}\varphi _{n+1}+j_n\varphi _n\right\} ,
\end{equation}
where $F^n\equiv \int d^3\mathbf{x}\mathcal{F}^n$ is the system function in
the absence of the source. There are other ways of introducing sources into
the system, but the above method was found to be most practical. Since these
sources are eventually switched off, it does not really matter how they are
introduced, as long as they are dealt with consistently according to the
principles of discrete time mechanics.

With the above system function the Cadzow equation of motion \cite
{J&N-II,CADZOW.70} is 
\begin{equation}
\frac \delta {\delta \varphi _n\left( \mathbf{x}\right) }\left\{
F^n+F^{n-1}\right\} +Tj_n\left( \mathbf{x}\right)  =
{}_{{}_{\!\!\!\!\!\!\!c}} \;\, 0,  \label{eqc}
\end{equation}
which reduces to 
\begin{equation}
\frac \partial {\partial \varphi _n}\left\{ \mathcal{F}^n+\mathcal{F}%
^{n-1}\right\} -\nabla \cdot \frac \partial {\partial \nabla \varphi
_n}\left\{ \mathcal{F}^n+\mathcal{F}^{n-1}\right\}
+Tj_n={}_{{}_{\!\!\!\!\!\!\!c}}\;\,0,
\end{equation}
where $\mathcal{F}^n$ is the system function density in the absence of
sources. Here we use the symbol $= {}_{{}_{\!\!\!\!\!\!\!c}}\ \thinspace$ to
denote an equality holding over a true or dynamical classical trajectory.

The action sum $A^{NM}\left[ j\right] $ in the presence of sources for
evolution between times $MT$ and $NT$ is 
\begin{eqnarray}
A^{NM}\left[ j\right] &=&A^{NM}+\frac{_1}{^2}T\int d^3\mathbf{x}\left\{
j_M\varphi _M+j_N\varphi _N\right\} \;\;  \nonumber \\
&&\;\;\;\;\;\;\;\;\;\;\;\;\;\;\;\;\;\;\;+T\sum_{n=M+1}^{N-1}\int d^3\mathbf{%
x\,}j_n\varphi _n,\;\;\;\;\;M<N.
\end{eqnarray}
Use of the discrete time Schwinger action principle \cite{J&N-I} 
\begin{equation}
\delta \langle \phi ,N|\psi ,M\rangle ^j=i\langle \phi ,N|\delta \hat{A}%
^{NM}\left[ j\right] |\psi ,M\rangle ^j,\;\;\;\;M<N
\end{equation}
leads to the functional derivatives 
\begin{eqnarray}
\frac{-i}T\frac \delta {\delta j_M\left( \mathbf{x}\right) }\langle \phi
,N|\psi ,M\rangle ^j &=&\frac{_1}{^2}\langle \phi ,N|\hat{\varphi}_M\left( 
\mathbf{x}\right) |\psi ,M\rangle ^j  \nonumber \\
\frac{-i}T\frac \delta {\delta j_n\left( \mathbf{x}\right) }\langle \phi
,N|\psi ,M\rangle ^j &=&\langle \phi ,N|\hat{\varphi}_n\left( \mathbf{x}%
\right) |\psi ,M\rangle ^j,\;\;\;\;M<n<N  \nonumber \\
\frac{-i}T\frac \delta {\delta j_N\left( \mathbf{x}\right) }\langle \phi
,N|\psi ,M\rangle ^j &=&\frac{_1}{^2}\langle \phi ,N|\hat{\varphi}_N\left( 
\mathbf{x}\right) |\psi ,M\rangle ^j.
\end{eqnarray}
Also, we find 
\begin{eqnarray}
\frac{-i}T\frac \delta {\delta j_n\left( \mathbf{x}\right) }\frac{-i}T\frac
\delta {\delta j_m\left( \mathbf{y}\right) }\langle \phi ,N|\psi ,M\rangle
^j &=&\langle \phi ,N|\hat{\varphi}_m\left( \mathbf{y}\right) \hat{\varphi}%
_n\left( \mathbf{x}\right) |\psi ,M\rangle ^j\left[ \Theta _{m-n}+\frac{_1}{%
^2}\delta _{m-n}\right]  \nonumber \\
&&+\langle \phi ,N|\hat{\varphi}_n\left( \mathbf{x}\right) \hat{\varphi}%
_m\left( \mathbf{y}\right) |\psi ,M\rangle ^j\left[ \Theta _{n-m}+\frac{_1}{%
^2}\delta _{m-n}\right] ,  \nonumber \\
&&\;\;\;\;\;\;\;\;\;\;\;\;\;\;\;\;\;\;\;\;\;\;\;(M<m,n<N)  \label{ord}
\end{eqnarray}
where $\Theta _n$ and $\delta _n$ are the discrete time step function and
discrete time delta defined in \emph{Paper} $I$. We will write (\ref{ord})
in the form 
\begin{equation}
\frac{-i}T\frac \delta {\delta j_n\left( \mathbf{x}\right) }\frac{-i}T\frac
\delta {\delta j_m\left( \mathbf{y}\right) }\langle \phi ,N|\psi ,M\rangle
^j=\langle \phi ,N|\tilde{T}\hat{\varphi}_n\left( \mathbf{x}\right) \hat{%
\varphi}_m\left( \mathbf{y}\right) |\psi ,M\rangle ^j,\;\;\;(M<m,n<N),
\end{equation}
where $\tilde{T}$ denotes discrete time ordering as discussed in \textit{%
Paper }$I$.

In applications we will normally be interested in the scattering limit $%
N=-M\rightarrow \infty $ and in matrix elements involving the \textit{in}
and \textit{out} vacua. We shall restrict our calculations to such matters.
This means we will discuss the $r$-point functions defined by 
\begin{eqnarray}
G_{n_1n_2...n_r}^j\left( \mathbf{x}_1,...,\mathbf{x}_r\right) &=&\frac{%
\langle 0^{out}|\tilde{T}\hat{\varphi}_{n_1}\left( \mathbf{x}_1\right) ...%
\hat{\varphi}_{n_r}\left( \mathbf{x}_r\right) |0^{in}\rangle ^j}{\langle
0^{out}|0^{in}\rangle ^j}  \nonumber  \label{n-point} \\
&=&\frac 1{Z\left[ j\right] }\frac{-i\delta }{T\delta j_{n_1}\left( \mathbf{x%
}_1\right) }...\frac{-i\delta }{T\delta j_{n_r}\left( \mathbf{x}_r\right) }%
Z\left[ j\right] ,  \label{dn-point}
\end{eqnarray}
where 
\begin{equation}
Z\left[ j\right] =\langle 0^{out}|0^{in}\rangle ^j
\end{equation}
is the ground state (vacuum) functional in the presence of the sources and $%
\tilde{T}$ denotes discrete time ordering.

An important question here concerns the existence of the ground state. In
common with most continuous time field theories, we have no general proof
that a ground state exists for interacting discrete time field theories.
Moreover, in discrete time mechanics there is no Hamiltonian as such, so the
question becomes more acute. However, for free fields, there will be what we
refer to as a compatible operator corresponding to some appropriate Logan
invariant \cite{J&N-I,LOGAN.73}. This is the nearest analogue to the
Hamiltonian in continuous time theory. Moreover, the appropriate compatible
operator for free neutral scalar fields is positive definite and this allows
a meaning for the \textit{in} and \textit{out} vacua to be given.

\section{The discrete time free scalar field}

\subsection{The free scalar field propagator}

Given the continuous time Lagrange density 
\begin{equation}
\mathcal{L}_0=\frac{_1}{^2}\partial _\mu \varphi \partial ^\mu \varphi -%
\frac{_1}{^2}\mu ^2\varphi ^2
\end{equation}
we use the virtual path approach discussed in \cite{J&N-II} to find the
system function density 
\begin{eqnarray}
\mathcal{F}_0^n &=&\frac{\left( \varphi _{n+1}-\varphi _n\right) ^2}{2T}%
-\frac T6\left( \nabla \varphi _{n+1}^2+\nabla \varphi _{n+1}\cdot \nabla
\varphi _n+\nabla \varphi _n^2\right)  \nonumber \\
&&-\frac{\mu ^2T}6\left( \varphi _{n+1}^2+\varphi _{n+1}\varphi _n+\varphi
_n^2\right) .  \label{11}
\end{eqnarray}
In the presence of the sources we take 
\begin{equation}
F_0^n\left[ j\right] =F_0^n+\frac{_1}{^2}T\int d^3\mathbf{x\,}\left\{
j_{n+1}\varphi _{n+1}+j_n\varphi _n\right\}
\end{equation}
and then the Cadzow equation of motion is 
\begin{equation}
\frac{\varphi _{n+1}-2\varphi _n+\varphi _{n-1}}{T^2}+\left( \mu ^2-\nabla
^2\right) \frac{\left( \varphi _{n+1}+4\varphi _n+\varphi _{n-1}\right) }6 =
{}_{{}_{\!\!\!\!\!\!\!c}} \;\, j_n.
\end{equation}
We now define the spatial Fourier transforms 
\begin{eqnarray}
\tilde{\varphi}_n\left( \mathbf{p}\right) &\equiv &\int d^3\mathbf{x\,}e^{-i%
\mathbf{p\cdot x}}\varphi _n\left( \mathbf{x}\right)  \nonumber \\
\tilde{j}_n\left( \mathbf{p}\right) &\equiv &\int d^3\mathbf{x\,}e^{-i%
\mathbf{p\cdot x}}j_n\left( \mathbf{x}\right) ,
\end{eqnarray}
and then the equation of motion becomes 
\begin{equation}
\left\{ \frac{\left( U_n-2+U_n^{-1}\right) }{T^2}+E^2\frac{\left(
U_n+4+U_n^{-1}\right) }6\right\} \tilde{\varphi}_n\left( \mathbf{p}\right) =
{}_{{}_{\!\!\!\!\!\!\!c}} \;\, \tilde{j}_n\left( \mathbf{p}\right) ,
\label{15}
\end{equation}
where $E\equiv \sqrt{\mathbf{p\cdot p+}\mu ^2}$ and $U_{n}$ is the
classical temporal displacement operator defined by 
\begin{equation}
U_nf_n\equiv f_{n+1}  \label{def}
\end{equation}
for any variable indexed by $n$. The solution to $\left( \ref{15}\right) $
with Feynman scattering boundary conditions is 
\begin{equation}
\tilde{\varphi}_n\left( \mathbf{p}\right) =\tilde{\varphi}_n^{\left(
0\right) }\left( \mathbf{p}\right) -T\sum_{m=-\infty }^\infty \tilde{\Delta}%
_F^{n-m}\left( \mathbf{p}\right) \tilde{j}_m\left( \mathbf{p}\right) ,
\end{equation}
where $\tilde{\varphi}_n^{\left( 0\right) }\left( \mathbf{p}\right) $ is a
solution to the homogeneous equation 
\begin{equation}
\left\{ \frac{\left( U_n-2+U_n^{-1}\right) }{T^2}+E^2\frac{\left(
U_n+4+U_n^{-1}\right) }6\right\} \tilde{\varphi}_n^{\left( 0\right) }\left( 
\mathbf{p}\right) =0
\end{equation}
and $\tilde{\Delta}_F^n\left( \mathbf{p}\right) $ is the discrete time
Feynman propagator in momentum space satisfying the equation 
\begin{equation}
\left\{ \frac{\left( U_n-2+U_n^{-1}\right) }{T^2}+E^2\frac{\left(
U_n+4+U_n^{-1}\right) }6\right\} \tilde{\Delta}_F^n\left( \mathbf{p}\right)
=-\frac{\delta _n}T.
\end{equation}
This equation for the propagator may be written in the form 
\begin{equation}
\left\{ U_n-2\eta _E+U_n^{-1}\right\} \tilde{\Delta}_F^n\left( \mathbf{p}%
\right) =-\Gamma _E\delta _n,
\end{equation}
where 
\begin{equation}
\Gamma _E=\frac{6T}{6+T^2E^2},\;\;\;\eta _E=\frac{6-2T^2E^2}{6+T^2E^2}.
\end{equation}

Using our experience with the discrete time harmonic oscillator propagator
discussed in \emph{Paper} $I$, we may immediately write down the solution
for the propagator in the form 
\begin{equation}
\tilde{\Delta}_F^n\left( \mathbf{p}\right) =\frac{\Gamma _E}{2i\sin \theta _E%
}e^{-i|n|\theta _E}=\frac{\Gamma _E}{2i\sin \theta _E}\left\{ e^{-in\theta
_E}\Theta _n+\delta _n+e^{in\theta _E}\Theta _{-n}\right\} ,  \label{form}
\end{equation}
where $\eta _E=\cos \theta _E.$ As discussed in \emph{Paper} $I,$ this
expression holds for the elliptic and the hyperbolic regimes with suitable
analytic continuation. In the continuous time limit $T\rightarrow
0,\,nT\rightarrow t,$ we recover the usual Feynman propagator in a spatially
Fourier transformed form, viz; 
\begin{eqnarray}
\lim_{_{T\rightarrow 0},_{\,n\rightarrow \infty ,\,nT=t}}\tilde{\Delta}%
_F^n\left( \mathbf{p}\right) &=&-\frac i{2E}\left\{ e^{-itE}\theta \left(
t\right) +e^{itE}\theta \left( -t\right) \right\}  \nonumber \\
&=&\int \frac{d\omega }{2\pi }\frac{e^{-i\omega t}}{\omega ^2-\mathbf{p\cdot
p}-\mu ^2+i\epsilon }  \nonumber \\
&=&\tilde{\Delta}_F\left( \mathbf{p},t\right) =\int d^3\mathbf{x\,}e^{-i%
\mathbf{p\cdot x}}\Delta _F\left( \mathbf{x},t\right) .
\end{eqnarray}

Turning to the quantisation process, the functional derivative satisfies the
rule 
\begin{equation}
\frac \delta {\delta j_n\left( \mathbf{x}\right) }j_m\left( \mathbf{y}%
\right) =\delta _{n-m}\delta ^3\left( \mathbf{x-y}\right) .
\end{equation}
With the definition 
\begin{equation}
\frac \delta {\delta \tilde{j}_n\left( \mathbf{p}\right) }\equiv \frac
1{\left( 2\pi \right) ^3}\int d^3\mathbf{x\,e}^{i\mathbf{p\cdot x}}\frac
\delta {\delta j_n\left( \mathbf{x}\right) }
\end{equation}
we find 
\begin{equation}
\frac \delta {\delta \tilde{j}_n\left( \mathbf{p}\right) }\tilde{j}_m\left( 
\mathbf{q}\right) =\delta _{n-m}\delta ^3\left( \mathbf{p-q}\right)
\end{equation}
and then 
\begin{equation}
\langle 0^{out}|\varphi _n\left( \mathbf{x}\right) |0^{in}\rangle ^j=-\frac
iT\frac \delta {\delta j_n\left( \mathbf{x}\right) }Z\left[ j\right]
\end{equation}
gives 
\begin{equation}
\langle 0^{out}|\tilde{\varphi}_n\left( \mathbf{p}\right) |0^{in}\rangle ^j=-%
\frac{i\left( 2\pi \right) ^3}T\frac \delta {\delta \tilde{j}_n\left( -%
\mathbf{p}\right) }Z\left[ j\right] .
\end{equation}
Hence we find 
\begin{equation}
Z_0\left[ j\right] =Z_0[0]\exp \left\{ -\frac{_1}{^2}iT^2\sum_{n,m=-\infty
}^\infty \int d^3\mathbf{x}d^3\mathbf{y\,}j_n\left( \mathbf{x}\right) \Delta
_F^{n-m}\left( \mathbf{x-y}\right) j_m\left( \mathbf{y}\right) \right\} ,
\label{Z0}
\end{equation}
where 
\begin{equation}
\Delta _F^n\left( \mathbf{x}\right) =\int \frac{d^3\mathbf{p}}{\left( 2\pi
\right) ^3}\,e^{i\mathbf{p\cdot x}}\tilde{\Delta}_F^n\left( \mathbf{p}%
\right) .
\end{equation}
This propagator satisfies the equation 
\begin{equation}
\left\{ \frac{U_n-2+U_n^{-1}}{T^2}+\left( \mu ^2-\nabla ^2\right) \frac{%
\left( U_n+4+U_n^{-1}\right) }6\right\} \Delta _F^n\left( \mathbf{x}\right)
=-\frac{\delta _n}T\delta ^3\left( \mathbf{x}\right) .
\end{equation}

\subsection{The free field commutators}

In this section we use (\ref{Z0}) to obtain the vacuum expectation value of
the free field commutators. Writing the propagator (\ref{form}) in the form 
\begin{equation}
\tilde{\Delta}_F^n\left( \mathbf{p}\right) =c_E\left[ z^{-n}\Theta _n+\delta
_n+z^n\Theta _{-n}\right]
\end{equation}
then we find for the elliptic and hyperbolic regimes 
\begin{eqnarray}
c_E &=&\frac{-i\sqrt{3}}{E\sqrt{12-T^2E^2}},\;\;\;TE<\sqrt{12}  \nonumber \\
&=&\frac{-3}{E\sqrt{T^2E^2-12}},\;\;\;TE>\sqrt{12.}
\end{eqnarray}
An application of (\ref{dn-point}) gives 
\begin{equation}
\langle 0|\hat{\varphi}_{n+1}\left( \mathbf{x}\right) \hat{\varphi}_n\left( 
\mathbf{y}\right) |0\rangle =i\Delta _F^1\left( \mathbf{x-y}\right) ,
\end{equation}
from which we deduce 
\begin{eqnarray}
\langle 0|\left[ \hat{\varphi}_{n+1}\left( \mathbf{x}\right) ,\hat{\varphi}%
_n\left( \mathbf{y}\right) \right] |0\rangle &=&-6iT\int \frac{d^3\mathbf{p}%
}{\left( 2\pi \right) ^3}\frac{e^{i\mathbf{p\cdot }\left( \mathbf{x-y}%
\right) }}{6+\left( \mathbf{p\cdot p}+\mu ^2\right) T^2}  \nonumber \\
&=&\frac{-6i}{4\pi T\left| \mathbf{x-y}\right| }e^{-\sqrt{\mu ^2+6/T^2}%
\left| \mathbf{x-y}\right| }.  \label{comm}
\end{eqnarray}
Both elliptic and hyperbolic regions of momentum space contribute to this
result, which has the form of a Yukawa potential function.

We turn now to the direct approach to quantisation discussed in \emph{Paper} 
$I$. If we define the momentum $\pi _n\left( \mathbf{x}\right) $ conjugate
to $\varphi _n\left( \mathbf{x}\right) $ via the rule $\pi _n\left( \mathbf{x%
}\right) \equiv -\frac \delta {\delta \varphi _n\left( \mathbf{x}\right)
}F^n $ then $\left( \ref{11}\right) $ gives 
\begin{equation}
\pi _n\equiv \frac{\varphi _{n+1}-\varphi _n}T+\frac T6\left( \mu ^2-\nabla
^2\right) \left( \varphi _{n+1}+2\varphi _n\right)
\end{equation}
for the free field. The naive canonical quantisation discussed in \emph{%
Paper }$I$ is equivalent in field theory terms to 
\begin{equation}
\left[ \hat{\pi}_n\left( \mathbf{x}\right) ,\hat{\varphi}_n\left( \mathbf{y}%
\right) \right] =-i\delta ^3\left( \mathbf{x-y}\right) ,
\end{equation}
from which we deduce 
\begin{eqnarray}
\left[ \hat{\varphi}_{n+1}\left( \mathbf{x}\right) ,\hat{\varphi}_n\left( 
\mathbf{y}\right) \right] &=&-6iT\int \frac{d^3\mathbf{p}}{\left( 2\pi
\right) ^3}\frac{e^{i\mathbf{p\cdot }\left( \mathbf{x-y}\right) }}{6+\left(
\mu ^2+\mathbf{p\cdot p}\right) T^2}  \nonumber \\
&=&\frac{-6i}{4\pi T\left| \mathbf{x-y}\right| }e^{-\sqrt{\mu ^2+6/T^2}%
\left| \mathbf{x-y}\right| },  \label{38}
\end{eqnarray}
assuming $\left[ \hat{\varphi}_n\left( \mathbf{x}\right) ,\hat{\varphi}%
_n\left( \mathbf{y}\right) \right] =0.$ This is consistent with the approach
to quantisation via the Schwinger action principle from which we obtained $%
\left( \ref{comm}\right) $.

We note that the reason this works is that the system function for a free
field is an example of what we call a \emph{normal system} in \emph{Paper} $%
I $. For interacting field theories this will no longer be the case and then
the commutators analogous to the above will probably no longer be
c-functions. We recall that in continuous time field theories, interacting
field commutators are not canonical in general either, so the analogies
between discrete and continuous time mechanics hold well here also.

For the free field particle creation and annihilation operators we define
the variables 
\begin{equation}
a_n\left( \mathbf{p}\right) =i\Gamma _E^{-1}e^{in\theta _E}\left[ \tilde{%
\varphi}_{n+1}\left( \mathbf{p}\right) -e^{i\theta _E}\tilde{\varphi}%
_n\left( \mathbf{p}\right) \right] ,
\end{equation}
where 
\begin{equation}
\Gamma _E=\frac{6T}{6+T^2E^2},\;\;\;\tilde{\varphi}_n\left( \mathbf{p}%
\right) =\int d^3\mathbf{x\,e}^{-i\mathbf{p\cdot x}}\varphi _n\left( \mathbf{%
x}\right)
\end{equation}
and the momentum $\mathbf{p}$ is restricted to the elliptic region. Then we
find 
\begin{equation}
\left[ \hat{a}_n\left( \mathbf{p}\right) ,\hat{a}_n^{+}\left( \mathbf{q}%
\right) \right] =2E\sqrt{1-T^2E^2/12}\,\left( 2\pi \right) ^3\delta ^3\left( 
\mathbf{p-q}\right)
\end{equation}
when we quantise and use $\left( \ref{38}\right) $. This tends to the
correct continuous time limit as $T\rightarrow 0$.

If we interpret the factor $2E\sqrt{1-T^2E^2/12}$ in the above as a particle
flux density then this will be indistinguishable from the conventional
density $2E$ in continuous time field theory for normal momenta, but falls
to zero as the parabolic barrier $TE=\sqrt{12}$ is approached from below.
This suggests that there is in principle a physical limit to the possibility
of creating extremely high momentum particle states in the laboratory or of
observing such particles in cosmic rays. This should have an effect on all
discussions involving momentum space, such as particle decay lifetime and
cross-section calculations, and in the long term, on unified field theories.

\section{Interacting Discrete Time Scalar Fields}

\subsection{Reduction formulae}

In applications to particle scattering theory we shall be interested in
incoming and outgoing physical particle states, with individual particle
energies satisfying the elliptic condition $TE<\sqrt{12}.$ We note that
energy is defined here via the linear momentum $\mathbf{p}$ by the rule 
\begin{equation}
E\equiv +\sqrt{\mathbf{p.p+}\mu ^2}.
\end{equation}
This is the \textit{only} meaning we give to the term \textit{energy}.

Given the annihilation and creation operators 
\begin{eqnarray}
\hat{a}_n\left( \mathbf{p}\right) &=&i\Gamma _E^{-1}\int d^3\mathbf{x\,}%
e^{in\theta _E-i\mathbf{p\cdot x}}\left[ \hat{\varphi}_{n+1}\left( \mathbf{x}%
\right) -e^{i\theta _E}\hat{\varphi}_n\left( \mathbf{x}\right) \right] 
\nonumber \\
\hat{a}_n^{+}\left( \mathbf{p}\right) &=&-i\Gamma _E^{-1}\int d^3\mathbf{x\,}%
e^{-in\theta _E+i\mathbf{p\cdot x}}\left[ \hat{\varphi}_{n+1}\left( \mathbf{x%
}\right) -e^{-i\theta _E}\hat{\varphi}_n\left( \mathbf{x}\right) \right]
\end{eqnarray}
where 
\begin{equation}
\Gamma _E=\frac{6T}{6+T^2E^2},\;\;\;\;\;\cos \theta _E\equiv \eta _E=\frac{%
6-2T^2E^2}{6+T^2E^2},  \label{theta}
\end{equation}
then a direct application of the standard reduction formalism gives the
reduced matrix elements 
\begin{eqnarray}
\langle \alpha ^{out}|\left( \tilde{T}\hat{\zeta}\right) \hat{a}%
_{in}^{+}\left( \mathbf{p}\right) |\beta ^{in}\rangle _R &=&i\sum_{n=-\infty
}^\infty \int d^3\mathbf{x}e^{-in\theta _E+i\mathbf{p\cdot x}}%
\overrightarrow{K_{n,\mathbf{p}}}\langle \alpha ^{out}|\tilde{T}\hat{\zeta}%
\hat{\varphi}_n\left( \mathbf{x}\right) |\beta ^{in}\rangle  \nonumber \\
&&
\end{eqnarray}
and 
\begin{eqnarray}
\langle \alpha ^{out}|\hat{a}_{out}\left( \mathbf{p}\right) \tilde{T}\hat{%
\zeta}|\beta ^{in}\rangle _R &=&i\sum_{n=-\infty }^\infty \int d^3\mathbf{x}%
e^{in\theta _E-i\mathbf{p\cdot x}}\overrightarrow{K_{n,\mathbf{p}}}\langle
\alpha ^{out}|\tilde{T}\hat{\zeta}\hat{\varphi}_n\left( \mathbf{x}\right)
|\beta ^{in}\rangle ,  \nonumber \\
&&
\end{eqnarray}
where $\hat{\zeta}$ denotes any collection of field operators and 
\begin{equation}
\overrightarrow{K_{n,\mathbf{p}}}\equiv \Gamma _E^{-1}\left( U_n-2\eta
_E+U_n^{-1}\right) .
\end{equation}
Using these results we can readily write down the scattering matrix for a
process consisting of $r$ incoming physical momentum particles with momenta $%
\mathbf{p}_1,\mathbf{p}_2,...,\mathbf{p}_r$ and $s$ out-going particles with
momenta $\mathbf{q}_1,\mathbf{q}_2,...,\mathbf{q}_s$.

\subsection{Interacting fields: scalar field theory}

We turn now to interacting scalar field theories based on continuous time
Lagrange densities of the form 
\begin{equation}
\mathcal{L=L}_0-V\left( \varphi \right) .
\end{equation}
In order to illustrate what happens in discrete time quantum field theory,
we shall discuss the details of a scalar field with a $\varphi ^3$
interaction term, deriving the analogue of the Feynman rules.

In the presence of sources the above Lagrange density leads to the system
function 
\begin{equation}
F^n\left[ j\right] =F_{\left( 0\right) }^n-T\int_0^1d\lambda \int d^3\mathbf{%
x\,}V\left( \tilde{\varphi}_n\right) +\frac{_1}{^2}T\int d^3\mathbf{x\,}%
\left\{ j_n\varphi _n+j_{n+1}\varphi _{n+1}\right\} ,
\end{equation}
where we use the virtual paths 
\begin{equation}
\tilde{\varphi}_n\left( \mathbf{x}\right) \equiv U_n^\lambda \varphi_n
\left( 
\mathbf{x}\right) =\lambda \varphi _{n+1}\left( \mathbf{x}\right) +\bar{%
\lambda}\varphi _n\left( \mathbf{x}\right) ,\;\;\;\;\;0\leq \lambda \leq
1,\;\;\;\bar{\lambda}\equiv 1-\lambda ,
\end{equation}
as discussed in \textit{Paper} $II$ for neutral scalar fields. Here and
below we shall find the operator 
\begin{equation}
U_n^\lambda \equiv \lambda U_n+\bar{\lambda}
\end{equation}
particularly useful, where $U_n$ is the classical temporal displacement
operator defined by $\left( \ref{def}\right) $. The vacuum functional is now
defined via the discrete time path integral 
\begin{eqnarray}
Z\left[ j\right] &=&\int \left[ d\varphi \right] \exp \left\{ iA\left[
j\right] \right\}  \nonumber \\
&\equiv &\prod_{n=-\infty }^\infty \left( \int_{\mathbf{x}}\left[ d\varphi
_n\right] \right) \exp \left\{ iA\left[ j\right] \right\} ,
\end{eqnarray}
where 
\begin{equation}
A\left[ j\right] \equiv \sum_{n=-\infty }^\infty F^n\left[ j\right]
=\sum_{n=-\infty }^\infty F_{\left( 0\right) }^n\left[ j\right] -i\Sigma
\!\!\!\!\!\!\int_{n\lambda \mathbf{x}}V\left( \tilde{\varphi}_n\right)
\end{equation}
and the $\varphi _{n}$are functionally integrated over their
spatially-indexed degrees of freedom. In the above and subsequently we shall
use the notation 
\begin{equation}
\Sigma \!\!\!\!\!\!\int_{n\lambda \mathbf{x}}\equiv T\int_0^1d\lambda
\sum_{n=-\infty }^\infty \int d^3\mathbf{x}
\end{equation}
whenever such a particular combination of spatial integration, summation,
and virtual path integration occurs. This replaces the four-dimensional
integral $\int d^4x\equiv \int dtd^3\mathbf{x}$ found in normal relativistic
field theory.

We now postulate our quantum dynamics to be governed by the equation 
\begin{equation}
\int \left[ d\varphi \right] \left[ \frac \delta {\varphi _n\left( \mathbf{x}%
\right) }\left\{ F^n+F^{n-1}\right\} +Tj_n\left( \mathbf{x}\right) \right]
\exp \left\{ iA\left[ j\right] \right\} =0,
\end{equation}
which is equivalent to a vacuum expectation value of the Heisenberg operator
equations of motion derived formally from Cadzow's equation $\left( \ref{eqc}%
\right) .$ Integrating by parts, we arrive at the more convenient expression 
\begin{equation}
Z\left[ j\right] =\exp \left\{ -i\Sigma \!\!\!\!\!\!\int_{n\lambda \mathbf{x}%
}V\left( \mathcal{D}_{n\lambda \mathbf{x}}\right) \right\} Z_0\left[
j\right] ,
\end{equation}
where 
\begin{eqnarray}
Z_0[j] &\equiv &\int \left[ d\varphi \right] \exp \left\{ iT\sum_{n=-\infty
}^\infty \int d^3\mathbf{x\,}\left( \mathcal{F}_0^n+j_n\varphi _n\right)
\right\}  \nonumber \\
&=&Z_0\left[ 0\right] \exp \left\{ -\frac{_1}{^2}iT^2\sum_{n,m=-\infty
}^\infty \int d^3\mathbf{x}d^3\mathbf{y\,}j_n\left( \mathbf{x}\right) \Delta
_F^{n-m}\left( \mathbf{x}-\mathbf{y}\right) j_m\left( \mathbf{y}\right)
\right\} \nonumber \\
\end{eqnarray}
and 
\begin{eqnarray}
\mathcal{D}_{n\lambda \mathbf{x}} &\equiv &\frac{-i}TU_n^\lambda \frac
\delta {\delta j_n\left( \mathbf{x}\right) }  \nonumber \\
&=&\frac{-i}T\left\{ \lambda \frac \delta {\delta j_{n+1}\left( \mathbf{x}%
\right) }+\bar{\lambda}\frac \delta {\delta j_n\left( \mathbf{x}\right)
}\right\} .
\end{eqnarray}

Turning now to $\varphi ^{3}$theory, we recall that with hindsight
the potential $V^{(3)}\left( \varphi \right) $ is normally taken to have the
form 
\begin{equation}
V^{\left( 3\right) }\left( \varphi \right) =\frac g{3!}\left\{ \varphi
^3-\Gamma \varphi \right\} ,
\end{equation}
where the (infinite) subtraction constant $\Gamma $ is formally given by 
\begin{equation}
\Gamma =3i\Delta _F\left( 0\right) .
\end{equation}
This has the role of cancelling off self-interaction loops at vertices in
the Feynman rules expansion programme. We find that for discrete time, the
same effect is achieved by taking the potential to have the form 
\begin{equation}
V^{\left( 3\right) }\left( \tilde{\varphi}_n\right) =\frac g{3!}\left\{ 
\tilde{\varphi}_n^3-\tilde{\Gamma}\tilde{\varphi}_n\right\}  \label{pot}
\end{equation}
where 
\begin{equation}
\tilde{\Gamma}=2i\Delta _F^0\left( \mathbf{0}\right) +\frac{_1}{^2}i\left[
\Delta _F^1\left( \mathbf{0}\right) +\Delta _F^{-1}\left( \mathbf{0}\right)
\right] .
\end{equation}

The first objective is to find a perturbative expansion for $Z\left[
j\right] ,$ which we write in the form 
\begin{equation}
Z\left[ j\right] =Z_0\left[ j\right] +Z_1\left[ j\right] +Z_2\left[ j\right]
+...
\end{equation}
where 
\begin{equation}
Z_p\left[ j\right] \equiv -\frac ip\Sigma \!\!\!\!\!\!\int_{n\lambda \mathbf{%
x}}V^{\left( 3\right) }\left( \mathcal{D}_{n\lambda \mathbf{x}}\right)
Z_{p-1}\left[ j\right] ,\;\;p=1,2,...
\end{equation}
Having found $Z\left[ j\right] $ we then calculate the required vacuum
expectation value of time ordered products of fields by functional
differentiation in the standard way. The results lead to a set of rules for
a diagrammatic expansion analogous to the Feynman rules in continuous time
theory, with specific differences. The details of the calculations are
omitted here as they are routine and tedious, but the results are as follows.

\subsection{Feynman rules for discrete time-ordered products}

The objective in this subsection is to present the rules for a diagrammatic
expansion of scattering amplitudes in the absence of external sources. The
latter are used merely to provide an internal handle on the correlation
functions of the theory and are set to zero at the end of the day. This
programme is carried out in two stages. In this subsection we give the rules
for the evaluation of successive terms in a Feynman diagram type of
expansion for the vacuum expectation value of the time ordered product 
\begin{equation}
\langle 0^{out}|\tilde{T}\hat{\varphi}_1\left( \mathbf{x}_1\right) \hat{%
\varphi}_2\left( \mathbf{x}_2\right) ...\hat{\varphi}_k\left( \mathbf{x}%
_k\right) |0^{in}\rangle
\end{equation}
with $k$ discrete time scalar fields; we shall give the rules for a system
with interaction given by $\left( \ref{pot}\right) $, so the expansion is
effectively in powers in the coupling constant $g$:

\begin{enumerate}
\item  first find the ordinary continuous time Feynman rules in space-time;

\item  draw all the different diagrams normally discussed in this programme;

\item  for a given diagram with $V$ vertices and $I$ internal lines find its
conventional weighting factor $\omega $, such as the well-known factor of $%
\frac{_1}{^2}$ for the simple loop in $\varphi ^{3}$ theory;

\item  at each vertex, associate a factor
\begin{equation}
igT\Sigma \!\!\!\!\!\!\int_{m\lambda \mathbf{z}}\equiv igT\int_0^1d\lambda
\sum_{m=-\infty }^\infty \int d^3\mathbf{z;}
\end{equation}

\item  for each external line running from the external point $\left( n,%
\mathbf{x}\right) $ to a vertex with indices $\left( m,\lambda ,\mathbf{z}%
\right) $ assign a propagator 
\begin{equation}
iU_m^\lambda \Delta _F^{m-n}\left( \mathbf{z-x}\right) ;
\end{equation}

\item  for each internal line running from vertex $\left( m_1,\lambda _1,%
\mathbf{z}_1\right) $ to vertex $\left( m_2,\lambda _2,\mathbf{z}_2\right) $
assign a propagator 
\begin{equation}
iU_{m_1}^{\lambda _1}U_{m_2}^{\lambda _2}\Delta _F^{m_2-m_1}\left( \mathbf{z}%
_2-\mathbf{z}_1\right) ;
\end{equation}

\item  do the $\lambda $ integrals.
\end{enumerate}

It is in general much more convenient to perform the virtual path
integrations (over the $\lambda $'s) after the diagrams have been written
down rather than before the diagrammatic expansion. In many cases the
operator $U_m^\lambda $ acting on an external propagator can be transferred
to act on internal propagators using the rule 
\begin{equation}
\sum_{m=-\infty }^\infty \left( U_m^\lambda f_m\right) g_m=\sum_{m=-\infty
}^\infty f_m\bar{U}_m^\lambda g_m,  \label{rule}
\end{equation}
for any indexed functions $f_n$, $g_{n},$where we define 
\begin{equation}
\bar{U}_m^\lambda g_m\equiv \lambda U_m^{-1}g_m+\bar{\lambda}g_m=\lambda
g_{m-1}+\bar{\lambda}g_m.
\end{equation}
However, this does not work so conveniently whenever two or more external
lines meet at the same vertex.

To illustrate these rules in operation consider the conventional
perturbation theory expansion of the time ordered product $\langle 0|T\hat{%
\varphi}\left( x_1\right) \hat{\varphi}\left( x_2\right) |0\rangle $ in
powers of the coupling constant. The conventional Feynman rules give the
expansion 
\begin{eqnarray}
\langle 0|T\hat{\varphi}\left( x_1\right) \hat{\varphi}\left( x_2\right)
|0\rangle &=&i\Delta _F\left( x_1-x_2\right)  \nonumber \\
&&-\frac{_1}{^2}g^2\int d^4z_1d^4z_2\,\Delta _F\left( x_1-z_1\right) \Delta
_F\left( z_1-z_2\right) \times  \nonumber \\
&&\Delta _F\left( z_2-z_1\right) \Delta _F\left( z_2-x_2\right) +O\left(
g^4\right) .
\end{eqnarray}
The second term on the right hand side corresponds to the single loop
diagram with $V=2$, $I=2$ in $\varphi ^{3}$ scalar theory and is
divergent. Part of the motivation for investigating discrete time field
theory is the hope that the corresponding diagram might be modified in some
significant way.

Using the rules outlined above the analogue expansion in discrete time gives 
\begin{eqnarray}
\langle 0|\tilde{T}\hat{\varphi}_{n_1}\left( \mathbf{x}_1\right) \hat{\varphi%
}_{n_2}\left( \mathbf{x}_2\right) |0\rangle &=&i\Delta _F^{n_1-n_2}\left( 
\mathbf{x}_1\mathbf{-x}_2\right)  \nonumber \\
&&-\frac{_1}{^2}g^2\Sigma \!\!\!\!\!\!\int_{m_1\lambda _1\mathbf{z}_1}\Sigma
\!\!\!\!\!\!\int_{m_2\lambda _2\mathbf{z}_2}\,\left\{ U_{m_1}^{\lambda
_1}\Delta _F^{m_1-n_1}\left( \mathbf{z}_1\mathbf{-x}_1\right) \right\} \times
\nonumber \\
&&\;\;\;\;\left\{ U_{m_1}^{\lambda _1}U_{m_2}^{\lambda _2}\Delta
_F^{m_2-m_1}\left( \mathbf{z}_2\mathbf{-z}_1\right) \right\} \left\{
U_{m_2}^{\lambda _2}U_{m_1}^{\lambda _1}\Delta _F^{m_1-m_2}\left( \mathbf{z}%
_1\mathbf{-z}_2\right) \right\} \times  \nonumber \\
&&\;\;\;\;\;\;\left\{ U_{m_2}^{\lambda _2}\Delta _F^{m_2-n_2}\left( \mathbf{z%
}_2\mathbf{-x}_2\right) \right\} +O\left( g^4\right) .
\end{eqnarray}

For this particular process the second term on the right hand side can be
rewritten using the rule $\left( \ref{rule}\right) $ to give 
\begin{eqnarray}
\langle 0|\tilde{T}\hat{\varphi}_{n_1}\left( \mathbf{x}_1\right) \hat{\varphi%
}_{n_2}\left( \mathbf{x}_2\right) |0\rangle &=&i\Delta _F^{n_1-n_2}\left( 
\mathbf{x}_1\mathbf{-x}_2\right)  \nonumber \\
&&-\frac{_1}{^2}g^2\Sigma \!\!\!\!\!\!\int_{m_1\lambda _1\mathbf{z}_1}\Sigma
\!\!\!\!\!\!\int_{m_2\lambda _2\mathbf{z}_2}\Delta _F^{m_1-n_1}\left( 
\mathbf{z}_1\mathbf{-x}_1\right) \times  \nonumber \\
&&\left\{ \bar{U}_{m_1}^{\lambda _1}\bar{U}_{m_2}^{\lambda _2}\left[
U_{m_1}^{\lambda _1}U_{m_2}^{\lambda _2}\Delta _F^{m_2-m_1}\left( \mathbf{z}%
_2\mathbf{-z}_1\right) \right] ^2\right\} \times  \nonumber \\
&&\Delta _F^{m_2-n_2}\left( \mathbf{z}_2\mathbf{-x}_2\right) +O\left(
g^4\right) ,
\end{eqnarray}
using the symmetry 
\begin{equation}
\Delta _F^n\left( \mathbf{x}\right) =\Delta _F^{-n}\left( -\mathbf{x}\right)
.
\end{equation}

The integrals over $\mu $ and $\lambda $ can be integrated at this stage to
give a multitude of subdiagrams distinguished by different split times,
which is the ultimate effect of the discretisation process. The various
subdiagrams contributing to the loop diagram are shown in $Figure\;1$, each
with a numerical factor. The sum over all numerical factors for this diagram
should add up to $144$. The full amplitude corresponding to the loop diagram
is the sum of each of these sub-diagrams, times the numerical factor for
each sub-diagram, divided by $288$, taking into account the original
weighting factor of one half. By using symmetry arguments it can be shown
that the twenty nine distinct diagrams in $Figure\;1$ reduce to the twelve
diagrams shown in $Figure\;2$.

The above rules are relevant to vacuum expectation values of discrete
time-ordered products of field operators. For particle scattering matrix
elements the rules become simpler, as discussed next.

\section{Scattering amplitudes}

We are now in a position to discuss particle scattering amplitudes. First we
explain how the scattering amplitude for a two-two particle scattering
process based on the box diagram, $Figure\;3,$ is calculated, and then we
shall state the results for the general scattering diagram. This diagram was
chosen because it involves a loop integration.

\subsection{The two-two box scattering diagram}

Consider two \emph{incoming} scalar particles with $3-$momenta $\mathbf{a}$, 
$\mathbf{b}$ respectively scattering via a the box scattering diagram shown
in \emph{Figure 3,} into two \emph{outgoing} particles with $3-$momenta $%
\mathbf{c}$ and $\mathbf{d}$ respectively. Each of these particles is
associated with a $\theta $ parameter as given by (\ref{theta}) which lies
in the physical particle interval $[0,\pi )$. Negative values of such a
parameter correspond to waves moving backwards in discrete time and would be
interpreted in the usual way as anti-particles in the Feynman-Stueckelberg
interpretation. Both positive and negative values occur in the discrete time
Feynman propagators, just as in conventional field theory.

Using the reduction formulae in $\S 4.1$ we may write for the scattering
reaction amplitude $S_{if}$%
\begin{eqnarray}
S_{if} &\equiv &\langle 0^{out}|\hat{a}_{out}\left( \mathbf{d}\right) \hat{a}%
_{out}\left( \mathbf{c}\right) \hat{a}_{in}^{+}\left( \mathbf{b}\right) \hat{%
a}_{in}^{+}\left( \mathbf{a}\right) |0^{in}\rangle _R  \nonumber \\
&=&i^4\left( \prod_{j=1}^4\sum_{n_j=-\infty }^\infty \int d^3\mathbf{x}%
_j\right) e^{-in_1\theta _{\mathbf{a}}+i\mathbf{a\cdot x}_1}e^{-in_2\theta _{%
\mathbf{b}}+i\mathbf{b\cdot x}_2}e^{in_3\theta _{\mathbf{c}}-i\mathbf{c\cdot
x}_3}e^{in_4\theta _{\mathbf{d}}-i\mathbf{d\cdot x}_4}\times  \nonumber \\
&&\overrightarrow{K_{n_1,\mathbf{a}}}\overrightarrow{K_{n_2,\mathbf{b}}}%
\overrightarrow{K_{n_3,\mathbf{c}}}\overrightarrow{K_{n_4,\mathbf{d}}}%
\langle 0^{out}|\tilde{T}\hat{\varphi}_{n_1}\left( \mathbf{x}_1\right) \hat{%
\varphi}_{n_2}\left( \mathbf{x}_2\right) \hat{\varphi}_{n_3}\left( \mathbf{x}%
_3\right) \hat{\varphi}_{n_4}\left( \mathbf{x}_4\right) |0^{in}\rangle ,
\nonumber \\
&&
\end{eqnarray}
where 
\begin{equation}
\overrightarrow{K_{n_1,\mathbf{a}}}\equiv \Gamma ^{-1}\left( \mathbf{a}%
\right) \left( U_{n_1}-2\eta \left( \mathbf{a}\right) +U_{n_1}^{-1}\right)
\end{equation}
with 
\begin{equation}
\Gamma \left( \mathbf{a}\right) \equiv \frac{6T}{6+T^2E_{\mathbf{a}}^2}%
,\;\;\;E_{\mathbf{a}}\equiv \sqrt{\mathbf{a\cdot a}+\mu ^2},\;\;\;\eta
\left( \mathbf{a}\right) \equiv \frac{6-2T^2E_{\mathbf{a}}^2}{6+T^2E_{%
\mathbf{a}}^2}=\cos \theta _{\mathbf{a}},
\end{equation}
and similarly for the other particles.

Next we expand the 4-point function according to the rules outlined in $\S
4.3$ and consider for the purposes of this discussion only the contribution
associated with the box diagram of \textit{Figure 3, }viz 
\begin{eqnarray}
&&\langle 0^{out}|\tilde{T}\hat{\varphi}_{n_1}\left( \mathbf{x}_1\right) 
\hat{\varphi}_{n_2}\left( \mathbf{x}_2\right) \hat{\varphi}_{n_3(}\mathbf{x}%
_3)\hat{\varphi}_{n_4}(\mathbf{x}_4)|0^{in}\rangle |_{BOX}  \nonumber \\
&&\;\;\;\;\;\;\;\;\;\;\;=\left( igT\right) ^4\left( \prod_{j=1}^4\Sigma
\!\!\!\!\!\!\int_{m_j\lambda _j\mathbf{z}_j}iU_{m_j}^{\lambda _j}\Delta
_F^{m_j-n_j}\left( \mathbf{z}_j-\mathbf{x}_j\right) \right) \times  \nonumber
\\
&&\;\;\;\;\;\;\;\;\;\;\;\;\;\;\;\;\;\left[ U_{m_2}^{\lambda
_2}U_{m_1}^{\lambda _1}\Delta _F^{m_2-m_1}\left( \mathbf{z}_2-\mathbf{z}%
_1\right) \right] \left[ U_{m_3}^{\lambda _3}U_{m_2}^{\lambda _2}\Delta
_F^{m_3-m_2}\left( \mathbf{z}_3-\mathbf{z}_2\right) \right] \times  \nonumber
\\
&&\;\;\;\;\;\;\;\;\;\;\;\;\;\;\;\;\;\;\;\;\left[ U_{m_4}^{\lambda
_4}U_{m_3}^{\lambda _3}\Delta _F^{m_4-m_3}\left( \mathbf{z}_4-\mathbf{z}%
_3\right) \right] \left[ U_{m_1}^{\lambda _1}U_{m_4}^{\lambda _4}\Delta
_F^{m_1-m_4}\left( \mathbf{z}_1-\mathbf{z}_4\right) \right] .  \nonumber \\
&&
\end{eqnarray}
The next step is to do the $\mathbf{x}_j$ integrals, converting the two
point functions on each external leg of the diagram to its momentum space
form, using 
\begin{equation}
\tilde{\Delta}_F^n\left( \mathbf{p}\right) =\int d\mathbf{x\,}e^{i\mathbf{%
p\cdot x}}\Delta _F^n\left( \mathbf{x}\right) .
\end{equation}
Then we use the result 
\begin{equation}
\overrightarrow{K_{n,\mathbf{p}}}\tilde{\Delta}_F^n\left( \mathbf{p}\right)
=-\delta _n,
\end{equation}
taking care to bring the operators and summations into the brackets whenever
the $U_m^\lambda $ operators occur. This effectively amputates the external
legs of the diagram. Then we can immediately carry out the summations over
the external integers $n_i$ and arrive at the simplified form 
\begin{eqnarray}
S_{if} &=&\left( gT\right) ^4\left( \prod_{j=1}^4\sum_{m_j=-\infty }^\infty
\int_0^1d\lambda _j\int d^3\mathbf{z}_j\right) e^{i\mathbf{a\cdot z}_1+i%
\mathbf{b\cdot z}_2-i\mathbf{c\cdot z}_3-i\mathbf{d\cdot z}_4}\times 
\nonumber \\
&&\left\{ U_{m_1}^{\lambda _1}e^{-im_1\theta _{\mathbf{a}}}\right\} \left\{
U_{m_2}^{\lambda _2}e^{-im_2\theta _{\mathbf{b}}}\right\} \left\{
U_{m_3}^{\lambda _3}e^{im_3\theta _{\mathbf{c}}}\right\} \left\{
U_{m_4}^{\lambda _4}e^{im_4\theta _{\mathbf{d}}}\right\} \times  \nonumber \\
&\;\;\;\;\;&\;\;\left[ U_{m_2}^{\lambda _2}U_{m_1}^{\lambda _1}\Delta
_F^{m_2-m_1}\left( \mathbf{z}_2\mathbf{-z}_1\right) \right] \left[
U_{m_3}^{\lambda _3}U_{m_2}^{\lambda _2}\Delta _F^{m_3-m_2}\left( \mathbf{z}%
_3-\mathbf{z}_2\right) \right] \times  \nonumber \\
&&\;\;\;\;\;\left[ U_{m_4}^{\lambda _4}U_{m_3}^{\lambda _3}\Delta
_F^{m_4-m_3}\left( \mathbf{z}_4-\mathbf{z}_3\right) \right] \left[
U_{m_1}^{\lambda _1}U_{m_4}^{\lambda _4}\Delta _F^{m_1-m_4}\left( \mathbf{z}%
_1-\mathbf{z}_4\right) \right] .
\nonumber \\
&&
\end{eqnarray}
Now we use the representation of the propagator 
\begin{equation}
\Delta _F^n\left( \mathbf{x}\right) \equiv \frac 1{\left( 2\pi \right)
^4}\int d^3\mathbf{k}\int_{-\pi }^\pi \frac{d\theta }Te^{-in\theta +i\mathbf{%
k\cdot x}}\tilde{\Delta}_F(\mathbf{k},\theta )
\end{equation}
and evaluate the $\mathbf{z}_{i}$integrals to find 
\begin{eqnarray}
&&S_{if}=g^4\left( 2\pi \right) ^3\delta ^3\left( \mathbf{a+b-c-d}\right)
\left( \prod_{j=1}^4\sum_{m_j=-\infty }^\infty \int_0^1d\lambda _j\frac
1{2\pi }\int_{-\pi }^\pi d\theta _j\right) \int \frac{d^3\mathbf{k}}{\left(
2\pi \right) ^3}\times  \nonumber \\
&&\;\;\;\;\;\;\left\{ U_{m_1}^{\lambda _1}e^{-im_1\theta _{\mathbf{a}%
}}\right\} \left\{ U_{m_2}^{\lambda _2}e^{-im_2\theta _{\mathbf{b}}}\right\}
\left\{ U_{m_3}^{\lambda _3}e^{im_3\theta _{\mathbf{c}}}\right\} \left\{
U_{m_4}^{\lambda _4}e^{im_4\theta _{\mathbf{d}}}\right\} \times  \nonumber \\
&&\;\;\;\;\;\;\left[ U_{m_2}^{\lambda _2}U_{m_1}^{\lambda
_1}e^{-i(m_2-m_1)\theta _1}\tilde{\Delta}_F^{}\left( \mathbf{k,}\theta
_1\right) \right] \left[ U_{m_3}^{\lambda _3}U_{m_2}^{\lambda
_2}e^{-i(m_3-m_2)}\tilde{\Delta}_F^{_{}}\left( \mathbf{k+b},\theta _2\right)
\right] \times  \nonumber \\
&&\;\;\;\;\;\;\;\;\left[ U_{m_4}^{\lambda _4}U_{m_3}^{\lambda
_3}e^{-i(m_4-m_3)}\tilde{\Delta}_F^{}\left( \mathbf{k+b-c,}\theta _3\right)
\right] \times  \nonumber \\
&&\;\;\;\;\;\;\;\;\;\;\left[ U_{m_1}^{\lambda _1}U_{m_4}^{\lambda
_4}e^{-i(m_1-m_4)}\tilde{\Delta}_F^{}\left( \mathbf{k-a,}\theta _4\right)
\right] .
\end{eqnarray}
Here we see the appearance of overall linear momentum conservation, as
expected. Next we use the result 
\begin{equation}
U_m^\lambda e^{im\theta }=e^{im\theta }f_\lambda \left( \theta \right)
\end{equation}
where 
\begin{equation}
f_\lambda \left( \theta \right) \equiv \lambda e^{i\theta }+\bar{\lambda},
\end{equation}
to find 
\begin{eqnarray}
S_{if} &=&g^4\left( 2\pi \right) ^3\delta ^3\left( \mathbf{a+b-c-d}\right)
\left( \prod_{j=1}^4\sum_{m_j=-\infty }^\infty \int_0^1d\lambda _j\frac
1{2\pi }\int_{-\pi }^\pi d\theta _j\right) \times  \nonumber \\
&&\int \frac{d^3\mathbf{k}}{\left( 2\pi \right) ^3}e^{-im_1\theta _{\mathbf{a%
}}}f_{\lambda _1}^{*}\left( \theta _a\right) e^{-im_2\theta _{\mathbf{b}%
}}f_{\lambda _2}^{*}\left( \theta _b\right) e^{im_3\theta _{\mathbf{c}%
}}f_{\lambda _3}\left( \theta _c\right) e^{im_4\theta _{\mathbf{d}%
}}f_{\lambda _4}(\theta _4)\times  \nonumber \\
&&\;\;\;\;\;e^{i(m_1-m_2)\theta _1}f_{\lambda _1}\left( \theta _1\right)
f_{\lambda _2}^{*}\left( \theta _1\right) \tilde{\Delta}_F^{}\left( \mathbf{%
k,}\theta _1\right)  \nonumber \\
&&\;\;\;\;\;\;\;e^{i(m_2-m_3)\theta _2}f_{\lambda _2}\left( \theta _2\right)
f_{\lambda _3}^{*}\left( \theta _2\right) \tilde{\Delta}_F^{_{}}\left( 
\mathbf{k+b},\theta _2\right) \times  \nonumber \\
&&\;\;\;\;\;\;\;\;\;\;e^{i(m_3-m_4)\theta _3}f_{\lambda _3}\left( \theta
_3\right) f_{\lambda _4}^{*}\left( \theta _3\right) \tilde{\Delta}%
_F^{}\left( \mathbf{k+b-c,}\theta _3\right) \times  \nonumber \\
&&\;\;\;\;\;\;\;\;\;\;\;\;\;e^{i(m_4-m_1)\theta _4}f_{\lambda _4}\left(
\theta _4\right) f_{\lambda _1}^{*}\left( \theta _4\right) \tilde{\Delta}%
_F^{}\left( \mathbf{k-a,}\theta _4\right) .  \nonumber \\
&&
\end{eqnarray}

We are now able to do the summations over the $m_{i}.$ We notice that
each summation gives a Fourier series representation of the periodic Dirac
delta, viz 
\begin{equation}
\sum_{m=-\infty }^\infty e^{imx}=2\pi \sum_{m=-\infty }^\infty \delta
(x+2m\pi )\equiv 2\pi \delta _P\left( x\right) ,
\end{equation}
and so we find 
\begin{eqnarray}
S_{if} &=&g^4\left( 2\pi \right) ^4\delta _P\left( \theta _a+\theta
_b-\theta _c-\theta _d\right) \delta ^3\left( \mathbf{a+b-c-d}\right) \left(
\prod_{j=1}^4\int_0^1d\lambda _j\right) \times  \nonumber \\
&&\int \frac{d^3\mathbf{k}}{\left( 2\pi \right) ^4}\int_{-\pi }^\pi d\theta
\,f_{\lambda _1}^{*}\left( \theta _a\right) f_{\lambda _2}^{*}\left( \theta
_b\right) f_{\lambda _3}\left( \theta _c\right) f_{\lambda _4}(\theta
_4)\times  \nonumber \\
&&f_{\lambda _1}\left( \theta \right) f_{\lambda _2}^{*}\left( \theta
\right) \tilde{\Delta}_F^{}\left( \mathbf{k,}\theta \right) f_{\lambda
_2}\left( \theta +\theta _b\right) f_{\lambda _3}^{*}\left( \theta +\theta
_b\right) \tilde{\Delta}_F^{_{}}\left( \mathbf{k+b},\theta +\theta _b\right)
\times  \nonumber \\
&&f_{\lambda _3}\left( \theta +\theta _b-\theta _c\right) f_{\lambda
_4}^{*}\left( \theta +\theta _b-\theta _c\right) \tilde{\Delta}_F^{}\left( 
\mathbf{k+b-c,}\theta +\theta _b-\theta _c\right) \times  \nonumber \\
&&f_{\lambda _4}\left( \theta -\theta _a\right) f_{\lambda _1}^{*}\left(
\theta -\theta _a\right) \tilde{\Delta}_F^{}\left( \mathbf{k-a,}\theta
-\theta _a\right) .  \nonumber \\
&&
\end{eqnarray}
The crucial significance of this step is that we see the appearance of a
conservation rule for the parameters $\theta $. This is despite the
non-existence of a Hamiltonian in our formulation and the fact that we have
not constructed a Logan invariant for the fully interacting system.

We may go further and do the $\lambda _{i}$integrals. We define the 
\textit{vertex function} 
\begin{eqnarray}
V\left( \theta _a,\theta _b\right) &\equiv &\int_0^1d\lambda \,f_\lambda
^{*}\left( \theta _{\mathbf{a}}\right) f_\lambda ^{*}\left( \theta _{\mathbf{%
b}}\right) f_\lambda \left( \theta _{\mathbf{a}}+\theta _{\mathbf{b}}\right)
\nonumber \\
&=&\frac{\cos \left( \theta _{\mathbf{a}}+\theta _{\mathbf{b}}\right) +\cos
\left( \theta _{\mathbf{a}}\right) +\cos \left( \theta _{\mathbf{b}}\right)
+3}6
\end{eqnarray}
and so get the final result 
\begin{eqnarray}
S_{if} &=&g^4\left( 2\pi \right) ^4\delta _P\left( \theta _a+\theta
_b-\theta _c-\theta _d\right) \delta ^3\left( \mathbf{a+b-c-d}\right) \times
\nonumber \\
&&\int \frac{d^3\mathbf{k}}{\left( 2\pi \right) ^4}\int_{-\pi }^\pi d\theta
\,V\left( \theta _a,-\theta \right) V\left( \theta _b,\theta \right) V\left(
\theta +\theta _b,-\theta _c\right) V\left( -\theta _d,\theta _a-\theta
\right) \times  \nonumber \\
&&\tilde{\Delta}_F^{}\left( \mathbf{k,}\theta \right) \tilde{\Delta}%
_F^{_{}}\left( \mathbf{k+b},\theta +\theta _b\right) \tilde{\Delta}%
_F^{}\left( \mathbf{k+b-c,}\theta +\theta _b-\theta _c\right) \tilde{\Delta}%
_F^{}\left( \mathbf{k-a,}\theta -\theta _a\right) .  \nonumber \\
&&  \label{final}
\end{eqnarray}
A diagrammatic representation of the above shows that $\theta $-conservation
occurs at every vertex.

\subsection{The vertex functions}

The vertex functions V$\left( \theta _{1,}\theta _2\right) $ represent a
degree of softening at each vertex arising from our temporal point splitting
via the system function. At each vertex the sum of the \textit{incoming} $%
\theta $ parameters is always zero, including inside loops, so the vertex
function always depends on two parameters only. If we had a $\varphi ^4$
interaction we expect the vertex function will depend on three parameters,
and so on. A graphical presentation of the vertex function is given in 
\textit{Figure }$4$\textit{. }The vertex function has a minimum value of one
quarter and attains its maximum value of unity when the $\theta $ parameters
are each zero. This corresponds to the continuous time limit $T\rightarrow 0$%
.

\subsection{The propagators}

The propagators used in the final amplitude ($\ref{final})$ are readily
found using the basic definition 
\begin{equation}
\tilde{\Delta}_F\left( \mathbf{p},\theta \right) \equiv T\sum_{n=-\infty
}^\infty e^{in\theta }\tilde{\Delta}_F^n\left( \mathbf{p}\right)
\end{equation}
and the equation 
\begin{equation}
\left( U_n-2\eta _E+U_n^{-1}\right) \tilde{\Delta}_F^n\left( \mathbf{p}%
\right) =-\Gamma _E\delta _n.
\end{equation}
Then 
\begin{equation}
2\left( \cos \theta -\eta _E\right) \tilde{\Delta}_F\left( \mathbf{p},\theta
\right) =-T\Gamma _E.
\end{equation}
Now we need to choose the correct solution for Feynman scattering boundary
conditions. This is done by referring to the Feynman $-i\epsilon $
prescription, which corresponds to the replacement of $E^2$ in the above by $%
E^2-i\epsilon .$ This in turn corresponds to the replacement 
\begin{equation}
\eta _E\rightarrow \eta _E+i\epsilon .
\end{equation}
Hence we arrive at the desired solution 
\begin{equation}
\tilde{\Delta}_F\left( \mathbf{p},\theta \right) =\frac{-T\Gamma _E}{2\left(
\cos \theta -\eta _E-i\epsilon \right) },  \label{fn}
\end{equation}
which holds for both the elliptic region $-1<\eta _E<1$ and for the
hyperbolic region $-2<\eta _E<-1.$ It may be verified that the indexed
propagators $\left( \ref{form}\right) $ are given by the integrals 
\begin{eqnarray}
\tilde{\Delta}_F^n(\mathbf{p)} &=&\frac 1{2\pi }\int_{-\theta }^\theta
d\theta \,e^{-in\theta }\frac{-\Gamma _E}{2\left( \cos \theta -\eta
_E-i\epsilon \right) }  \nonumber \\
&=&\frac{\Gamma _E}{2\pi i}\oint \frac{dz}{z^n\left( z^2-2(\eta _E+i\epsilon
)z+1\right) },
\end{eqnarray}
the contour of integration being the unit circle in the anticlockwise sense.
We find for example 
\begin{equation}
\tilde{\Delta}_F^n\left( \mathbf{p}\right) =\frac{\Gamma _E}{2i\sin \theta _E%
}e^{-i|n|\theta _E}
\end{equation}
in the case of the elliptic regime, $T^2E^2<12,$ and 
\begin{equation}
\tilde{\Delta}_F^n\left( \mathbf{p}\right) =\frac{\left( -1\right)
^{n+1}\Gamma _E}{2\sinh \gamma _E}e^{-|n|\gamma _E}
\end{equation}
in the hyperbolic regime, $T^2E^2>12$. Here we make the parametrisation 
\begin{equation}
\cos \left( \zeta \right) \equiv \eta _E=\frac{6-2T^2E^2}{6+T^2E^2},
\end{equation}
where $\zeta $ is a complex parameter running just below the real axis from
the origin to $\pi $ (when $\zeta $ is written as $\theta _E)$ and then from 
$\pi $ to $\pi -i\ln (2+\sqrt{3})$ (when $\zeta $ is written in the form $%
\pi -i\gamma _E).$

If in $\left( \ref{fn}\right) \;$we introduce the variable $p_0$ related to $%
\theta $ by the rule 
\begin{equation}
\cos \theta \equiv \frac{6-2p_0^2T^2}{6+p_0^2T^2},\;\;\;\;sign\left( \theta
\right) =sign\left( p_0\right) ,  \label{cos}
\end{equation}
then we find 
\begin{equation}
\tilde{\Delta}_F\left( \mathbf{p,}\theta \right) =\frac 1{p_0^2-\mathbf{p}%
^2-m^2+i\epsilon }+\frac{T^2p_0^2}{6\left( p_0^2-\mathbf{p}^2-m^2+i\epsilon
\right) },  \label{ex}
\end{equation}
an exact result. From this we see the emergence of Lorentz symmetry as an
approximate symmetry of the mechanics. If $p_0$ in the above is taken to
represent the zeroth component of a four-vector, with the components of $%
\mathbf{p}$ representing the remaining components, then we readily see that
the first term on the right-hand side of $\left( \ref{ex}\right) $ is
Lorentz invariant. The second term is not Lorentz invariant, but we note it
is proportional to $T^2.$ If, as we expect, $T$ represents an extremely
small scale, such as the Planck time or less, then it is clear that Lorentz
symmetry should emerge as an extremely good approximate symmetry of our
mechanics.

\subsection{Comments}

The significance of our results is that not only is spatial momentum
conserved during a scattering process, as expected from the Maeda-Noether 
\cite{J&N-I,MAEDA.81} theorem , but the sum of the $\theta $ parameters of
the incoming particles is conserved. This is the discrete time analogue of
energy conservation, since in the limit $T\rightarrow 0$ we note 
\begin{equation}
\lim_{T\rightarrow 0}\frac{\theta _{\mathbf{p}}}T=E_{\mathbf{p}}=\sqrt{%
\mathbf{p\cdot p}+\mu ^2}.
\end{equation}

The $\theta $ conservation rule is unexpected at first sight in that we have
not discussed as yet any Logan invariant for the full interacting system
function. It appears that the analogue of energy conservation occurs here
because of the way in which we have set up our incoming and outgoing states
and allowed the scattering process to take place over infinite time. The
result would probably not hold for scattering over a finite time intervals,
which would be the analogue of the time-energy uncertainty relation in
conventional quantum theory. In essence, the LSZ scattering postulates
relate the Logan invariant for \emph{in-}states to the Logan invariant for
the \emph{out}-states in such a way that knowledge of the Logan invariant
for the intermediate time appears not to be required. This is true of the
scattering formalism as we have demonstrated, but the bound state question
would be a different matter.

Although the conservation of $\theta -$parameters during scattering
processes comes as a surprise it is a welcome one. Before the calculations
were done explicitly, it was believed that the energy conservation rule in
continuous time scattering processes would only arise in the limit $%
T\rightarrow 0$. Such a phenomenon was discussed by Lee \cite{LEE.83} in his
discrete time mechanics, which differs from ours in that his time intervals
are determined by the dynamics. That there is an exact conservation rule for 
\textit{something} in our discrete time scattering processes regardless of
the magnitude of $T$ is an indicator of the existence of some Logan
invariant. The surprise is that the something turns out to be the sum of the
incoming $\theta $ parameters, which suggests that our parametrisation of
the harmonic oscillator discussed in \textit{Paper }$I$ was a fortuitously
good one.

We point out here that our parameter $\theta $ is really an angle, unlike
conventional energy $E$, or $p_0$ in the above, and there is an implied
periodicity. However, because there is no concept of Hamiltonian or energy
in our theory, this periodicity does not cause any physically relevant side
effects. Incoming or outgoing particles will be \textit{on-shell} in the
sense that their associated $\theta $ parameter can be restricted to takes a
value in the interval $[0,\pi )$. Given this, then we may invert $\left( \ref
{cos}\right) $ to find 
\begin{equation}
\theta =p_0T-\frac 1{24}p_0^3T^3+O\left( T^5\right) .
\end{equation}

Another welcome feature is the modification of the propagators and the
appearance of vertex softening in the scattering diagrams. A detailed
discussion of the effects these features have on the divergences of various
loop integrals found in the conventional Feynman diagram programme will be
reserved for a subsequent paper.

The scattering amplitude found above for $Figure\;3$ reduces to the correct
continuous time amplitude in the limit $T\rightarrow 0$.

\subsection{Rules for scattering amplitudes}

We are now in a position to use our experience with the box diagram \textit{%
Figure }$3$ to write down the general rules for scattering diagrams.
Consider a scattering process with $a$ incoming particles with momenta $%
\mathbf{p}_1,\mathbf{p}_2,...,\mathbf{p}_a$ respectively, and $b$ outgoing
particles with momenta $\mathbf{q}_1,\mathbf{q}_2,...\mathbf{q}_b$
respectively. Make a diagrammatic expansion in the traditional manner of
Feynman. For each diagram do the following:

\begin{enumerate}
\item  at each vertex, conserve linear momentum and $\theta $ parameters,
i.e., the algebraic sum of \textit{incoming }momenta is zero and the
algebraic sum of the \textit{incoming} $\theta $ parameters is zero;

\item  at each vertex associate a factor 
\begin{equation}
igTV\left( \theta _1,\theta _2\right) ,
\end{equation}
where $\theta _1$ and $\theta _2$ are any two of the three \textit{incoming} 
$\theta $ parameters;

\item  for each internal line carrying momentum $\mathbf{k}$ and $\theta $
parameter, associate a factor 
\begin{equation}
iT^{-1}\tilde{\Delta}_F^{}\left( \mathbf{k,}\theta \right) ;
\end{equation}

\item  for each loop integral, a factor 
\begin{equation}
\int \frac{d^3\mathbf{k}}{\left( 2\pi \right) ^4}\int_{-\pi }^\pi d\theta ;
\end{equation}

\item  an overall momentum-$\theta $ parameter conservation factor 
\begin{equation}
\left( 2\pi \right) ^4\delta _P\left( \theta _{p_1}+...+\theta _{pa}-\theta
_{q_1}-...-\theta _{qb}\right) \delta ^3\left( \mathbf{p}_1+\mathbf{...+p}_a-%
\mathbf{q}_1-...-\mathbf{q}_b\right) 
\end{equation}

\item  a weight factor $\omega $ for each diagram, exactly as for the
standard Feynman rules.
\end{enumerate}

\section{Examples}

We are now in a position to give a number of examples of scattering
amplitude calculations using the above rules. We restrict our attention to $%
\varphi ^{3}$ theory as an illustrative example. QED and the
associated discrete time Feynman rules will be the subject of
\textit{Paper }$VI$ in this series.

\subsection{Figure 5a:}

Consider the basic single vertex diagram of \emph{figure 5a }with particles
with linear momentum $\mathbf{a,\;b}$ fusing to form a particle with linear
momentum $\mathbf{c}$. Overlooking the fact that this process gives zero for
on-shell momenta our discrete time Feynman rules give 
\begin{eqnarray}
S_{5a} &=&igT\left( 2\pi \right) ^4\delta _P\left( \theta _a+\theta
_b-\theta _c\right) \delta ^3\left( \mathbf{a+b-c}\right) V\left( \mathbf{a,b%
}\right) .  \nonumber \\
&&
\end{eqnarray}

\subsection{Figure 5b}

This diagram has a single loop. We find 
\begin{eqnarray}
S_{5b} &=&-g^3\left( 2\pi \right) ^4\delta _P\left( \theta _a+\theta
_b-\theta _c\right) \delta ^3\left( \mathbf{a+b-c}\right) \int \frac{d^3%
\mathbf{k}}{\left( 2\pi \right) ^4}\int_{-\pi }^\pi d\theta \times  \nonumber
\\
&&\,V\left( \theta _a,-\theta \right) V\left( \theta _b,\theta -\theta
_c\right) V\left( -\theta _c,\theta \right) \,\tilde{\Delta}_F\left( \mathbf{%
k},\theta \right) \tilde{\Delta}_F\left( \mathbf{k-c},\theta -\theta
_c\right) \tilde{\Delta}_F\left( \mathbf{k-a,}\theta -\theta _a\right) . 
\nonumber \\
&&
\end{eqnarray}
The nature of this diagram will be discussed in detail in subsequent papers.

\subsection{Figures 5c,d,e\thinspace \thinspace}

The order $g^2$ two-two scattering diagrams \emph{figures 5c,d,e }give 
\begin{eqnarray}
S_{5cde} &=&-g^2T\left( 2\pi \right) ^4\delta _P\left( \theta _a+\theta
_b-\theta _c-\theta _d\right) \delta ^3\left( \mathbf{a+b-c-d}\right) \times
\nonumber \\
&&\;\;\;\left\{ V\left( \theta _a\mathbf{,-}\theta _c\right) V\left( \theta
_b\mathbf{,-}\theta _d\right) \tilde{\Delta}_F^{}\left( \mathbf{a-c,}\theta
_a-\theta _c\right) \right. +  \nonumber \\
&&\;\;\;\;\;\;\;V\left( \theta _a\mathbf{,}\theta _b\right) V\left( \mathbf{-%
}\theta _c\mathbf{,-}\theta _d\right) \tilde{\Delta}_F^{}\left( \mathbf{a+b,}%
\theta _a+\theta _b\right) +  \nonumber \\
&&\;\;\;\;\;\;\;\;\;\;\;\left. V\left( \theta _a\mathbf{,-}\theta _d\right)
V\left( \theta _b\mathbf{,-}\theta _c\right) \tilde{\Delta}_F^{}\left( 
\mathbf{a}-\mathbf{d,}\theta _a-\theta _d\right) \right\} .  \nonumber \\
&&
\end{eqnarray}

\subsection{Figure 5f}

This diagram is an example of a higher order tree diagram process involving
no loops. We find 
\begin{eqnarray}
S_{5f} &=&ig^3T\left( 2\pi \right) ^4\delta _P\left( \theta _a+\theta
_b-\theta _c-\theta _d-\theta _e\right) \delta ^3\left( \mathbf{a+b-c-d-e}%
\right) \times  \nonumber \\
&&\;\;\;V\left( \theta _a\mathbf{,-}\theta _c\right) V\left( \theta _b%
\mathbf{,-}\theta _d\right) V\left( \theta _a\mathbf{-}\theta _c\mathbf{,}%
\theta _b\mathbf{-}\theta _d\right) \times  \nonumber \\
&&\;\;\;\;\;\;\;\tilde{\Delta}_F^{}\left( \mathbf{a-c},\theta _a-\theta
_c\right) \tilde{\Delta}_F^{}\left( \mathbf{d-b},\theta _d-\theta _b\right) .
\end{eqnarray}

\subsection{Figure 5g}

This diagram has a simple propagator loop and gives 
\begin{eqnarray}
S_{5g} &=&\frac{_1}{^2}g^4\left( 2\pi \right) ^4\delta _P\left( \theta
_a+\theta _b-\theta _c-\theta _d\right) \delta ^3\left( \mathbf{a+b-c-d}%
\right) \times  \nonumber \\
&&\tilde{\Delta}_F\left( \mathbf{a+b},\theta _a+\theta _b\right) \left\{
\frac 1{\left( 2\pi \right) ^4}\int d^3\mathbf{k}\int_{-\theta }^\theta
d\theta \,V\left( \theta _a,\theta _b\right) V\left( \theta _a+\theta
_b,-\theta \right) \right. \times  \nonumber \\
&&\left. V\left( \theta ,-\theta _a-\theta _b\right) V\left( -\theta
_c-\theta _d\right) \tilde{\Delta}_F\left( \mathbf{k},\theta \right) \tilde{%
\Delta}\left( \mathbf{k-a-b},\theta -\theta _c-\theta _d\right) \right\}
\times  \nonumber \\
&&\;\;\;\;\;\tilde{\Delta}_F\left( \mathbf{a+b},\theta _a+\theta _b\right) .
\end{eqnarray}
The question of the divergence of this integral will be reserved for a
subsequent paper.

\section{Concluding remarks}

The application of the principles outlines in \emph{Papers} $I$ and $II$ to
scalar field theory has indicated that the conventional programme of
constructing Feynman rules for scattering amplitudes goes over well into
discrete time. Of course there are differences, and it is to be hoped that
some of these will alleviate if not overcome some of the divergence problems
of the conventional field theory programme. An important point is that there
occurs in our approach a natural scale provided by $T$. It is possible that
this will provide a renormalisation cutoff scale which will not have to be
introduced by hand. Issues of renormalisation and divergence will be
discussed in a later paper.

A particularly important result which was not anticipated before the
diagrams were calculated is the conservation of the total $\theta $
parameters over a scattering process. This occurs even though no Logan
invariant corresponding to the total Hamiltonian has been found for the
fully interacting theory.

An important point to consider is the question of relativistic covariance.
Clearly our process of temporal discretisation breaks Lorentz covariance,
and with it the Poincar\'{e} algebra. However, it should be admitted by any
critic that there is actually no empirical evidence that Special Relativity
holds all the way up to infinite momentum. It is only an abstraction from
limited experience that it does. Therefore, the Poincar\'{e} algebra has no
more than the status of a really useful synthesis of limited experience. By
requiring our parameter $T$ to be small enough we should be able to
reproduce all of the good predictions of continuous time mechanics, with the
possibility of alleviating, if not removing, those aspects which are known
to cause problems, such as divergences in the renormalisation programme.
Moreover, we have given a principle based on the cosmic background radiation
field for finding a unique local inertial frame in which time is discretised.

Finally, if our discrete time programme could be caught out in a fatal way,
then we would have what amounts to a proof that time is really continuous.
This in itself makes our investigation a worthwhile one.

\section{Acknowledgement}

Keith Norton is grateful to the Crowther Fund of the Open University for
financial assistance during this course of this research.

\pagebreak

\end{document}